\documentclass[sn-mathphys-num]{sn-jnl}
\usepackage{graphicx, bm}%
\usepackage{multirow}%
\usepackage{amsmath,amssymb,amsfonts}%
\usepackage{amsthm}%
\usepackage{mathrsfs}%
\usepackage[title]{appendix}%
\usepackage{xcolor}%
\usepackage{textcomp}%
\usepackage{manyfoot}%
\usepackage{booktabs}%
\usepackage{algorithm}%
\usepackage{algorithmicx}%
\usepackage{algpseudocode}%
\usepackage{listings}%

\newcommand{\be}{\begin{equation}}
\newcommand{\ee}{\end{equation}}
\newcommand{\ba}{\begin{eqnarray}}
\newcommand{\ea}{\end{eqnarray}}

\begin{document}

\title[Exploring pulsar glitches with dipolar supersolids]{Exploring pulsar glitches with dipolar supersolids}

\author[1]{\fnm{Thomas} \sur{Bland}}\email{Thomas.Bland@uibk.ac.at}
\equalcont{These authors contributed equally to this work.}

\author*[1,2]{\fnm{Francesca} \sur{Ferlaino}}\email{Francesca.Ferlaino@uibk.ac.at}
\equalcont{These authors contributed equally to this work.}

\author*[3]{\fnm{Massimo} \sur{Mannarelli}}\email{Massimo.Mannarelli@lngs.infn.it}
\equalcont{These authors contributed equally to this work.}

\author[1]{\fnm{Elena} \sur{Poli}}\email{Elena.Poli@uibk.ac.at}
\equalcont{These authors contributed equally to this work.}

\author[3,4]{\fnm{Silvia} \sur{Trabucco}}\email{Silvia.Trabucco@gssi.it}
\equalcont{These authors contributed equally to this work.}


\affil[1]{\orgname{Universit\"{a}t Innsbruck, Fakult\"{a}t f\"{u}r Mathematik, Informatik und Physik}, \orgdiv{Institut f\"{u}r Experimentalphysik}, \postcode{6020} \city{Innsbruck}, \country{Austria}}

\affil[2]{\orgdiv{Institut f\"{u}r Quantenoptik und Quanteninformation}, \orgname{Osterreichische Akademie der Wissenschaften}, \orgaddress{\street{Technikerstraße 21a}, \city{Innsbruck}, \postcode{6020},  \country{Austria}}}

\affil[3]{\orgdiv{Laboratori Nazionali del Gran Sasso}, \orgname{INFN}, \orgaddress{\street{Via G.Acitelli, 22}, \city{Assergi (AQ)}, \postcode{I-67100},  \country{Italy}}}

\affil[4]{\orgdiv{Gran Sasso Science Institute}, \orgaddress{\street{ Viale Francesco Crispi, 7}, \city{L'Aquila}, \postcode{I-67100},  \country{Italy}}}

\abstract{
Glitches are sudden spin-up events that interrupt the gradual spin-down of rotating neutron stars. They are believed to arise from the rapid unpinning of vortices in the neutron star inner crust. The analogy between the inner crust of neutron stars and dipolar supersolids allows to  investigate glitches. Employing such analogy, we numerically analyze the vortex trapping mechanism and how the matter density distribution influences glitches. These results pave the way for the quantum simulation of celestial bodies in laboratories.}


\maketitle

\section{Introduction}\label{sec:introduction}
In physics, analogies are extremely powerful tools. They give the opportunity to look at physical phenomena from different perspectives, favoring connections between very disparate research fields. In some cases, appropriate analogies can be used to explore phenomena happening in seemingly unreachable objects, such as neutron stars (NSs). In recent work, we established  an intriguing  analogy between the inner crust of NSs and ultracold dipolar supersolids and used it  to investigate the anomalies in the rotation frequency of NSs, known as glitches\,\cite{Poli:2023vyp}.

Neutron stars are among the most mysterious and extreme objects populating our universe. Their observed  mass $\sim 1-2 M_\odot $, 
where $M_\odot$
is the solar mass, is compressed by the gravitational attraction into a radius of  about $10$ km. 
They are believed to form  after the gravitational collapse of massive progenitor stars in so-called supernova type II events\,\cite{Shapiro:1983du, Glendenning:1997wn}.  After a very short time period, on the order of  seconds,  they cool down and reach a stable configuration with gravitational pull balanced by the combined effect of degenerate Fermi pressure and repulsive short-range strong interaction. Inside NSs the matter density increases with radial depth, determining an
onion-like equilibrium structure\,\cite{Haensel:1993zw, Lattimer:2000nx, Douchin:2001sv, Haensel:2007yy, Potekhin:2013qqa, Sharma:2015bna, Blaschke:2018mqw, FiorellaBurgio:2018dga}. The outer crust is a solid material consisting of a lattice of neutron-rich nuclei. As the density increases in the inner crust, nuclei are packed so tightly that neutrons start dripping out, overlapping with each other, and forming a superfluid background. The neutron superfluid density distribution is not uniform but  partially retains the same crystalline order of the underlying protons\,\cite{Negele:1971vb}. From this perspective, the inner crust of a NS is simultaneously solid and superfluid. Finally, the core of NSs consists  of a homogeneous superfluid of neutrons, protons and electrons. Muons, heavy hadrons and more exotic particles may be present as well\,\cite{Shapiro:1983du, Glendenning:1997wn, Haensel:1993zw}.

In the majority of cases, NSs are observed as pulsars: regularly rotating emitters of electromagnetic radiation.  The first documented signs of these elusive celestial bodies were periodic radio frequency flashes discovered in 1967, resembling a galactic lighthouse\,\cite{1968Natur.217..709H,Hamil:2015hqa, Kaspi:2016jkv, Zhou:2022cyp}, so perfect that some interpreted it as an attempt at alien communication.  The lighthouse effect  results from the misalignment between the rotation and magnetization axes leading to a secular loss of rotational energy with a corresponding slow  decrease of the pulsar rotation frequency, $\Omega$. Apart from this steady spinning down, it has been observed that the rotation frequency of  pulsars occasionally shows anomalous jumps--called ``glitches"--in the form of an abrupt speed-up of the pulsar rotation followed by a slow relaxation close to its original value. It is precisely the observations of such pulsar glitches that have provided the first evidence of superfluidity in neutron-star interiors~\cite{BAYM1969n, Ruderman:1972aj, Pines1991,Link:1999ca, Haskell:2015jra, Zhou:2022cyp}.

Glitches can be seen as the consequence of the trapping and releasing of  vortices within a crystalline structure in the interior of NSs. However, unlike superfluids observed on Earth, which contain tens or hundreds of vortices, NSs  are expected to generate up to $\sim$10$^{17}$ vortices, posing a nontrivial question regarding how the results obtained with few vortex systems could be scaled. Due to the underlying crystalline structure, superfluid vortices cannot move freely: they are forced to position themselves between the crystal sites\,\cite{Anderson:1975, Alpar:1977, Alparcreep1, Alparcreep2, Link:1992mdl}. When the radiation emission slows down the outer crust, the vortices trapped in the inner crust fix their speed, and they do not allow the neutron superfluid component to slow down in parallel with the outer crust and any stellar component that is electromagnetically tightly coupled. Eventually, this unbalanced situation cannot be maintained; an array of vortices escapes from the inner crust, transferring angular momentum to the outer crust. Then, the outer crust spins up, resulting in a glitch. Although the sketched mechanism is  plausible, there is no direct evidence that it is the right one. 

In\,\cite{Poli:2023vyp} we explored whether such phenomenon can be reproduced using ultracold atoms. 
The field of ultracold quantum gases has proven to be an ideal tool for simulating various quantum phenomena, as it allows exceptionally precise experimental control over external conditions and internal interactions~\cite{Altman:2019vbv, tsakadze1973measurement, Tsakadze:1980, Warszawski:2011vy, Warszawski:2012ns, Warszawski:2012wa, graber2017neutron,  You:2024rtl, Alana:2024ziy, Magierski:2024gvu}. In this field, an object exhibiting both superfluid and crystalline properties, known as a ``supersolid"\,\cite{Leggett1970cas}, is well-known\,\cite{Li2017asp,Leonard2017sfi,Boettcher2019tsp,Tanzi2019ooa,Chomaz2019lla,norcia2021tds,Bland2022tds}.  Recently,  supersolids were realized in experiments with ultracold dipolar atoms\,\cite{Boettcher2019tsp,Tanzi2019ooa,Chomaz2019lla}, behaving like small magnets. Atoms such as erbium and dysprosium, when cooled to temperatures near absolute zero, organize into a liquid with both a crystalline structure and superfluid properties. This first experimental observation occurred simultaneously in 2019 in three different laboratories in Innsbruck, Pisa, and Stuttgart. 
The supersolid phase was observed with crystallization in 2D \cite{norcia2021tds,Bland2022tds}, and  recently quantum vortices were observed in dipolar superfluids\,\cite{Klaus2022} using magnetostirring. The same technique was employed to produce vortices in  supersolids\, 
\cite{Casotti:2024xns}. 
These 
results indicate that we might soon expect the experimental observation  of glitches in rotating ultracold atoms.

The present paper is organized as follows. In Sec.\,\ref{sec:methods} we briefly review how the NS  inner crust evolution can be simulated with ultracold atoms.  The vortex structure and dynamics are described in Sec.\,\ref{sec:vortex} employing the so-called imprinting procedure. We draw our conclusions in Sec.\,\ref{sec:conclusion}.

\section{Methods}\label{sec:methods}
We consider an ultracold gas of $^{164}$Dy bosons confined in an optical trap. Appropriately tuning the s-wave interaction $a_s$, it is possible to drive the system into a supersolid phase\,\cite{Recati:2023urk}. Rotating 2D supersolids can host quantized vortices\,\cite{Roccuzzo2020ras,Gallemi2020qvi,Ancilotto2021vpi, Casotti:2024xns}, thus allowing us to investigate the vortex dynamics underlying pulsar glitches.

In the numerical simulation, we prepare an initial state rotating with constant angular velocity, then we apply an external torque to mimic the pulsar's secular loss of angular velocity.
The time evolution of the system is determined by two coupled equations: the angular momentum balance equation, taking into account that an external torque is applied to the system, and the extended Gross-Pitaevskii equation, which determines the evolution of the macroscopic wavefunction. 

For both dipolar supersolids and NSs, the total angular momentum can be cast as
\be\label{eq:Ldecomp}
L_\text{tot}=L_\text{s} + L_\text{vort}\,,
\ee
where $L_\text{s}$ is the solid component, and $L_\text{vort}$ is the sum of the  angular momenta of vortices. 
In supersolids, one has access to the total angular momentum, as the expectation value of the system angular momentum, and the solid-part angular momentum. Thus, we  use   the above expression to obtain $ L_\text{vort}$. Notice that $ L_\text{vort}$ cannot be numerically evaluated counting the number of vortices, as is typically done in homogeneous superfluids, because the  vortex angular momentum depends on its position in the trap\,\cite{butts:1999, Ancilotto2021vpi}. 
The solid-part angular momentum is given by
\be\label{eq:Ls}
L_\text{s} = I_\text{s} \Omega\,,
\ee
where $I_\text{s}$ is the solid moment of inertia. This quantity should not be confused with the moment of inertia $I_{\text{mass}}$  determined by the matter distribution.
They differ because a supersolid  does not in general respond to an applied external torque as a  rigid body, hence 
$I_\text{s}/I_\text{mass}\leq 1$. The mass and supersolid moment of inertia coincide only when all the atoms form  droplets; in this case 
the whole mass of the system rigidly responds to an external torque. We model the $I_\text{s}$ component as
\begin{equation}
  I_\text{s}(t) = (1-f_\text{NCRI})I_\text{mass}(t)\,,  
  \label{eq:Is}
\end{equation}
where the non-classical rotational inertia coefficient, $f_\text{NCRI}$,
 captures the reduced rotational response of the system\,\cite{Leggett1970cas}.  For a homogeneous superfluid $f_\text{NCRI}=1$, while $f_\text{NCRI}=0$ for a rigid body.
Assuming that this quantity is not strongly dependent on the rotation frequency, we approximate 
 \be f_\text{NCRI}=1-\frac{I_\text{s,0}}{I_\text{mass,0}}\,, \ee
where $I_\text{s,0}$  and $I_\text{mass,0}$ are the values extrapolated in the limit of vanishing rotation\,\cite{Leggett1970cas}.

In NSs, the torque originates from the  electromagnetic emission of radiation that tends to slowly spin down the star. Such an effect is mimicked in dipolar supersolids by externally ramping down the trap rotation frequency. The angular momentum evolution of both NSs and supersolids can then be written as
\be
\dot L_\text{s} = -N_\text{em} - \dot{L}_\text{vort}\,, 
\ee
where $N_\text{em}$ is the external torque and we used the angular momentum decomposition in Eq.\,\eqref{eq:Ldecomp}. 
From Eq.\,\eqref{eq:Ls} we obtain that
\begin{align} \label{eq:glitchmodel}
I_\text{s}\dot\Omega = -N_\text{em} - \dot{L}_\text{vort} - \dot{I}_\text{s}\Omega\,,
\end{align} 
where $\dot{I}_\text{s}$ characterizes the variation of the solid component inertia.
In  NSs' glitches,  this final term is assumed to be negligible. In any case, it cannot be directly linked to any astrophysical observable, so far. In supersolids, we can trace the full system evolution, hence we evaluate it using Eq.\,\eqref{eq:Is}. Assuming constant $f_\text{NCRI}$,  
\be \label{eq:Is_dot}
\dot I_\text{s} =  (1-f_\text{NCRI}) \dot I_\text{mass}(t)\,,\ee 
where the mass moment of inertia 
for a rotation along the $\hat{z}$--axis
can be evaluated  from the standard definition: 
\be I_\text{mass}=\langle x^2+y^2\rangle\,.\ee
The expectation value is computed using the  macroscopic wavefunction $\Psi(\mathbf{r}, t)$:   
the solution of the extended Gross-Pitaevskii equation (eGPE)\,\cite{Waechtler2016qfi,Bisset2016gsp,FerrierBarbut2016ooq,Chomaz2016qfd}, 
\begin{align}\label{eq:eGPE}
i \hbar \frac{\partial \Psi}{\partial t} = (1-i\gamma)\left[\mathcal{L}[\Psi;a_s,a_\text{dd},\bm{\omega}] - \Omega\hat{L}_z\right]\Psi\,.
\end{align}
where $\mathcal{L}$ is the eGPE operator  and $\hat{L}_z  = x\hat{p}_y – y\hat{p}_x$ is the appropriate  angular momentum operator. The interparticle interactions depend on the s-wave scattering length, $a_s$, and the dipole-dipole scattering length,  $a_\text{dd}$, see\,\cite{Poli:2023vyp} for more details. 
 The dipole-dipole interaction in  $^{164}$Dy is such that $a_\text{dd} \simeq 130.8a_0$,  where $a_0$ is the Bohr radius. 

The evolution is not unitary since a small dissipation parameter, $\gamma$, is included. In this way, we can impart a rotation in the system when the trap is being rotated.
For $\gamma =0$ a supersolid in a circularly symmetric trap would not respond to any change in the rotation frequency.  We present here the results of numerical simulations for $\gamma=0.05$, however changing this value in a reasonable range does not strongly affect  our results\,\cite{Poli:2023vyp}.   Dissipation implies that the total energy is not conserved,  mimicking the fact that in realistic systems  part of the energy is transformed in heat. 
On the other hand, the total atom number of particles is conserved, thus we assume negligible particle leaks. We have taken constant $N = \int \text{d}^3\textbf{r}|\Psi|^2 = 3\times10^5$, which serves to properly normalize the macroscopic wavefunction, and trap frequencies $\omega_r=2\pi\times 50$ Hz and    $\omega_z = 2\pi \times 130$ Hz. 

\begin{figure}[t!]
\centering
\includegraphics[width=0.48\textwidth]{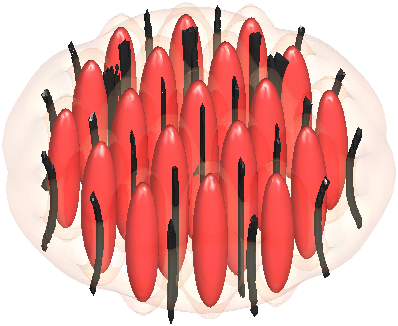}
\caption{Frame of the three-dimensional simulation of the rotating supersolid obtained with $a_s = 91 a_0$ and initial rotation frequency $\Omega (0) = 0.5 \omega_r$, see the text for more details. The red clusters indicate high number density. The combined effect of the dipole-dipole interaction and s-wave repulsion favors the formation of superfluid clusters.  The black lines are the vortex cores. Vortices are trapped in between the high density clusters. Close to the boundary, vortices are not straight lines: they bend to follow the curvature of the optical trap.}\label{fig:frame}
\end{figure}

 The evolution of the system is obtained by recursively solving Eqs.\,\eqref{eq:glitchmodel} and \eqref{eq:eGPE}.
We begin with solving  the  eGPE  with initial conditions given in imaginary time, at fixed $\Omega(0) = 0.5\omega_r$. This provides the initial wavefunction. Then, we evaluate $L_\text{tot} = \langle \hat{L}_z\rangle$ and  the superfluid angular momentum is obtained from Eq.\,\eqref{eq:Ldecomp}, namely $L_\text{vort} = L_\text{tot} - L_\text{s}$. From the mass distribution we determine the mass momentum of inertia and the variation of the  solid moment of inertia is computed using Eq.\,\eqref{eq:Is_dot}. The updated rotation frequency is then evaluated by means of Eq.\,\eqref{eq:glitchmodel} and it is  fed  in the eGPE, which  is solved to provide the updated wavefunction. Then the cycle starts over. 
In Fig.\,\ref{fig:frame} we report a frame of the three-dimensional simulation of the rotating supersolid. The red clusters correspond to high number density, they result from the interplay between the contact s-wave repulsion and the long range dipole-dipole interaction.
Vortices (black lines), are  found in the low density regions, between the clusters and close to the trap boundary.

\begin{figure}[t!]
\centering
\includegraphics[width=0.6\textwidth]{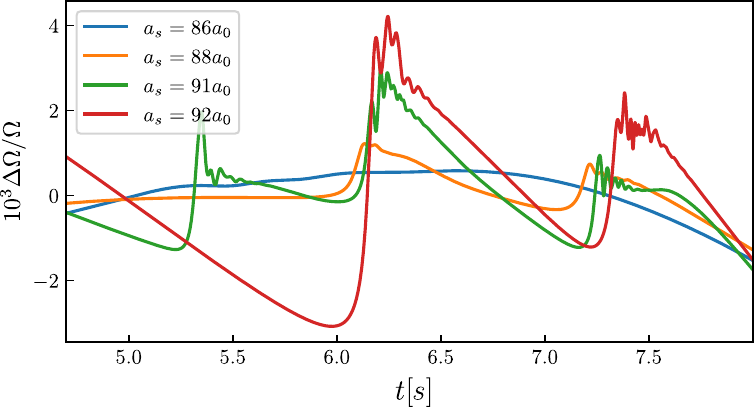} 
 \caption{Supersolid glitches for different values of the s-wave scattering length, $a_s$.
The blue line is obtained with  $a_s = 86 a_0$,  the orange line with  $a_s = 88 a_0$, the green line with $a_s = 91 a_0$ and the red line with $a_s = 92 a_0$.}
\label{fig:glitches}
\end{figure}

Fig.\,\ref{fig:glitches} shows that with increasing values of $a_s$  the   rotation frequency jump during a glitch becomes bigger\,\cite{Poli:2023vyp},  while for $a_s = 86 a_0$, glitches are not observed: the rotation frequency smoothly changes in time. The reason of this behavior is that for  $a_s = 86 a_0$ the superfluid component between droplets almost completely vanishes. With increasing  $a_s$ the interdroplet superfluid increases and vortices carry a larger fraction of angular momentum.  When the trap spins down these vortices  cannot adiabatically migrate towards the trap surface: they are pinned by the crystal.  The unpinning process is not smooth: a vortex unpins only when the lag between the solid and the vortex components is sufficiently large. As a consequence, during a glitch there is a sudden and relatively large  angular momentum ejection due to the unpinned vortices, which is transferred to the droplets, see Eq.\,\eqref{eq:glitchmodel}.

\section{Analyzing the vortex structure and dynamics}\label{sec:vortex}
A semi-quantitative description of the vortex structure and dynamics can be obtained by  imprinting   a vortex line in a specific position of a non-rotating supersolid. This allows us to  understand how a vortex modifies the matter distribution, as well as to estimate the vortex energy cost. In the imprinting procedure the  total wavefunction is approximated  multiplying the ground state wavefunction $\Psi_0(\textbf{r})$ of a nonrotating supersolid by the ansatz wave function 
\be\label{eq:imprinting}
    \Phi_v(x,y;x_0,y_0) =  \frac{\sqrt{(x-x_0)^2+(y-y_0)^2}}{\sqrt{(x-x_0)^2+(y-y_0)^2 + \Lambda^{-2}}} e^{i\theta}\,,
    \ee
that  imprints a vortex with phase $\theta = \arctan(y/x)$ on the top of the non-rotating background. Here $\Lambda$ is a regulator\,\cite{Bradley2012eso}  and ($x_0,y_0$) is the vortex center. Since $\Phi_v$ is  independent of the $z$--coordinate, the vortex  extends as a straight line along the $z$--direction.

Actually, in rotating dipolar supersolids the phase profile is not an azimuthal 2$\pi$ winding, but rather a complex pattern modified by the underlying crystal structure\,\cite{Ancilotto2021vpi}. Moreover, the finite size of the trap determines the bending of vortex lines close to the trap boundary, see Fig.\,\ref{fig:frame}.  However, the  simple ansatz that the total wavefunction is \be\Psi(\textbf{r};x_0,y_0) = \Psi_0(\textbf{r})\Phi_v(x,y;x_0,y_0) \,,\label{eq:vortexansatz}\ee  with $\bm r = (x,y)$, captures the main features of the  expected force field felt by vortices in a rotating trap\,\cite{Poli:2023vyp}, in particular close to the trap center where the vortex lines are straight. This is due to the fact, confirmed by numerical simulations, that rotation  does not strongly perturb  the droplet structure.

\begin{figure}[t]
\centering
\includegraphics[width=1\textwidth]{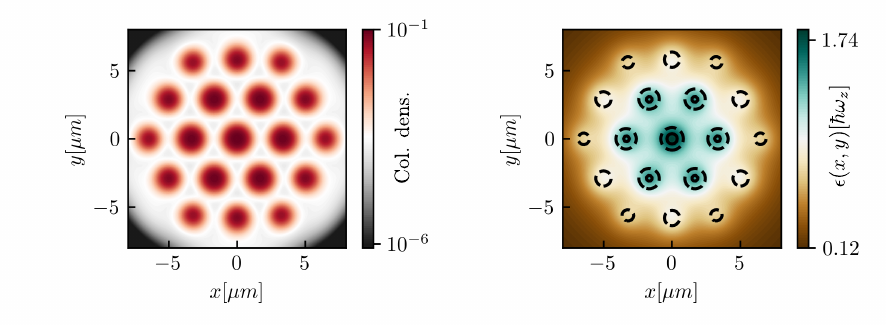} 
\caption{Energy cost of vortex seeding. Left panel: column density in $\mu$m$^{-2}$.  High density regions correspond to superfluid clusters self-organized in a nearly-triangular crystalline lattice.
Right panel: energy cost of imprinting a vortex to a non-rotating dipolar supersolid, in units of $\hbar \omega_z$. The black lines show the $30\%$ (dashed) and $85\%$ (solid) contours of the maximum column density. The energy  peaks where the number density is larger. Both plots are obtained  
with $N=3\times10^5$ bosons, trap frequency $\bm{\omega} = (\omega_r,\omega_z) = 2\pi\times(50,130)$Hz, and scattering lengths $a_\text{dd} = 130.8a_0$ and $a_s=90 a_0$, see the main text for more details. }\label{fig:density_energy}
\end{figure}


The way in which the matter  distribution shapes the vortex structure can be inferred from Fig.\,\ref{fig:density_energy}.
The left panel shows the column density profile of the  nonrotating supersolid  (obtained integrating the total density along the $z$--direction). Peaks of the column density correspond to  clusters of $^{164}$Dy bosons.
 In the region between 
clusters the  density is much smaller.
The right panel of Fig.\, \ref{fig:density_energy} shows  the 
energy cost  imprinting a vortex at any given position. It is obtained by  evaluating the energy  with the wavefunction given in Eq.\,\eqref{eq:vortexansatz}, over the full spatial range.  Note that the vortex energy cost peaks  coincide with the number density peaks: vortices prefer to sit in low-density regions.

\begin{figure}[t]
\centering
\includegraphics[scale=0.8]{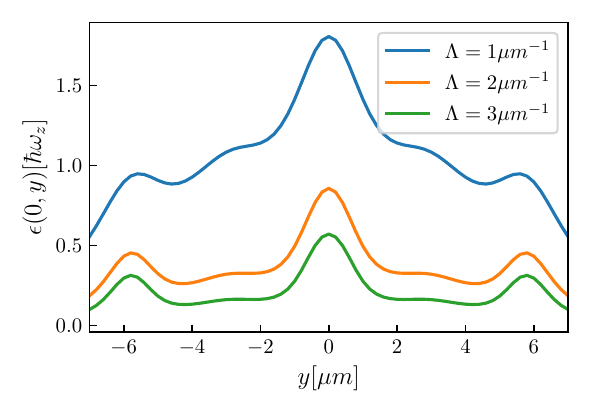}
\caption{Energy cost of imprinting a vortex in a nonrotating supersolid at $x=0$ varying $y$ for three different values of $\Lambda$, see Eq.\,\eqref{eq:imprinting}. 
The peak at $x=0, y=0$, coincides with the number density absolute  maximum: it is the most unfavorable position for a vortex. The local maxima at $y\simeq 6 \mu$m  coincide with the density local maxima. In between the local density maxima, for $\Lambda=2\,\mu$m$^{-1}$ and $3\,\mu$m$^{-1}$, there are two minima and a local maximum. The local maximum corresponds to a saddle point, while the two minima are the pinning sites. From the energy difference between the first minimum and the saddle point, one can estimate the unpinning energy barrier $\simeq 10^{-3} \hbar \omega_z$.   For $\Lambda=1$   such energy barrier tends to vanish.
} 
\label{fig:vortex_energy}
\end{figure}

In Fig.\,\ref{fig:vortex_energy} we report the vortex energy cost  at  $x=0$ varying $y$, for different values of $\Lambda$.
The central peak  coincides with  the central cluster of bosons.  
It is the absolute maximum: this is the least favorable position to place a vortex.  The other local maxima correspond to the  nearest clusters, while  minima correspond to the pinning sites. Starting from the centre and increasing $|y|$,  the energy cost  decreases until it reaches a local minimum. In other words, a vortex trying to move from $x=0, y=0$, along the $y$--direction feels a  potential well at
$y\simeq 1.8\,\mu$m, and then a  small barrier given by the difference between the energy at the local  maximum at $y\simeq 3\,\mu$m (actually a saddle point)  and the energy at the local  minimum at
$y\simeq 1.8\,\mu$m. Such energy barrier is  of the order of $10^{-3} \hbar\omega_z$. 
Note that this  barrier tends to vanish for $\Lambda = 1 \mu$m$^{-1}$. In the real time simulation, we  observe  vortex pinning in these local minima, which however may  differ in numerical simulations with systems of smaller size\,\cite{Alana:2024ziy}.



Regarding the vortex dynamics, 
the real time numerical simulations developed in\,\cite{Poli:2023vyp} allowed us to follow the vortex  starting from the ground state. 
In a rotating trap, all vortices are initially pinned in  local minima of the density. When the trap's rotation velocity is ramped down, one or more vortices may suddenly unpin along the more energetically favorable direction\,\cite{Liu2024}.    Useful information on the vortex motion  can be obtained by the imprinting procedure. In particular, one can imprint a vortex in an energetically unstable point and then follow its evolution.  At distances from the vortex core less than $\Lambda^{-1}$,  
 imprinted vortices can be treated as  point-like objects\,\cite{PhysRev.151.100}.  Whether this is a good approximation is hard to judge, since  in supersolids the healing length is not well-defined, see however the discussion in\,\cite{Gallemi2020qvi}. 

\begin{figure}[t!]
\centering
\includegraphics[width=0.5\textwidth]{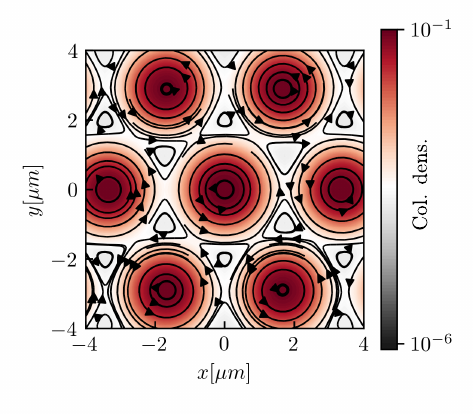}\includegraphics[width=0.5\textwidth]{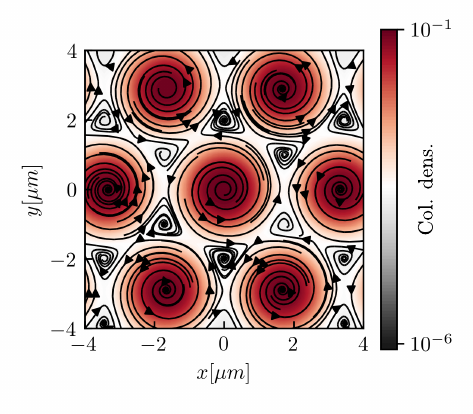}
\caption{Streamlines of the vortex velocity field obtained using Eq.~\eqref{eq:v_vortex_approx} with the  column number density (in units $\mu$m$^{-2}$) obtained for a non-rotating supersolid with $a_s= 90 a_0$. Left panel is  for vanishing dissipation, $\gamma =0$, 
the vortex velocity is tangent to the isodensity lines. 
Right panel includes dissipation, with $\gamma =0.05$. In this case, there is  a radial velocity component that drives the vortex to the local minima of the number density.}
\label{fig:velocities}
\end{figure}

In any case, within such approximation,  the vortex velocity follows from the general expression valid in an inhomogeneous Bose-Einstein condensates\,\cite{PhysRevLett.72.2426, PhysRevA.97.023617,  bland2023vortices}
\be\label{eq:v_vortex}
\bm{v}_v =  \frac{\hbar}{m}\left\{\nabla(\Phi - \gamma \log\sqrt \rho)  - \bm{\hat z} \times \nabla(\log\sqrt \rho+ \gamma \Phi) \right\} + \bm \Omega \times \bm r\,,
\ee
that depends  on the
 density and phase of the condensate in the absence of the vortex. 
 In principle, all contributions to  
the vortex velocity  may be of comparable importance\,\cite{PhysRevA.97.023617}. However, in the supersolid ground state in the absence of vortices, the phase is homogeneous, while space gradients of the density are large. Thus, in the corotating frame we  approximate
\be\label{eq:v_vortex_approx}
\bm{v}_v \simeq  -\frac{\hbar}{m}\left\{\gamma\nabla  \log\sqrt \rho  + \bm{\hat z} \times \nabla\log\sqrt \rho \right\} = \bm{v}_\perp + \bm{v}_\parallel \,,
\ee
consisting of two components: $\bm{v}_\perp$  and $\bm{v}_\parallel$,  orthogonal and tangential to the isodensity surfaces, respectively. Note that the $\bm{v}_\perp$ component, which pushes vortices away from high density regions,  is proportional to the dissipation parameter. In the absence of dissipation the vortex cannot seek the most energetically favorable position, its energy is frozen.  Indeed, for $\gamma=0$ the system is conservative: from Eq.\,\eqref{eq:v_vortex_approx}, the vortex velocity components can be written as
\be
\dot x_v =  -\frac{\hbar}{2m} \partial_x \log \rho\,, \quad
\dot y_v =  \frac{\hbar}{2m} \partial_y \log \rho\,,
\ee
and thus $d \log \rho/dt  =0 $, meaning that along an  orbit $\rho$ is constant. Since isodensity lines are  closed curves,  vortices follow periodic stable  orbits. 
There are however exceptions. Orbits consisting of  minima and maxima of the density are just points. Moreover, orbits including a saddle point are not periodic: they terminate at the saddle point. 
Regarding the  periodic orbits, we distinguish those  around density maxima from those around density minima. Including dissipation, see Eq.\,\eqref{eq:v_vortex_approx}, the former develop an unstable focus, while the latter have an asymptotically stable focus.

\begin{figure}[t!]
\centering
\includegraphics[width=0.5\textwidth]{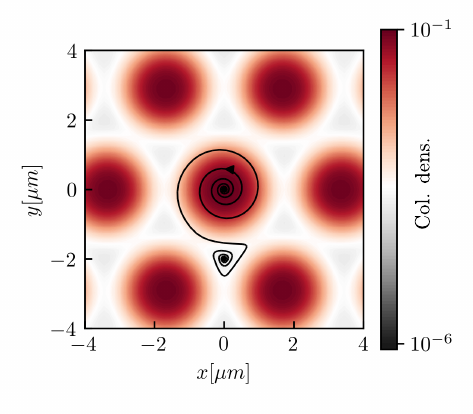}
\caption{Vortex orbit obtained using Eq.~\eqref{eq:v_vortex_approx} for a nonrotating supersolid with $\gamma =0.05$ and $a_s = 90a_0$. The  column number density is in units $\mu$m$^{-2}$. The vortex spirals out  the high density region toward a density minimum.}
\label{fig:trajectory}
\end{figure}

Figure \,\ref{fig:velocities} shows the streamlines  obtained using the column number density, $\rho(x,y)$,  of a nonrotating supersolid. Different column  densities correspond to different colors while the arrows represent  the vortex velocity. For vanishing dissipation ($\gamma =0$), left panel of Fig.\,\ref{fig:velocities}, the vortex velocity is always tangent to the isodensity lines. 
Vortices close to the droplet centre  move
following  almost circular  lines. They rotate around the centre of the droplet, pretty much as vortices rotate along circles corresponding to isodensity lines of  a Bose-Einstein condensate confined in an axially symmetric optical trap. 
The effect of dissipation is  shown in the right panel of Fig.\,\ref{fig:velocities}. The vortex velocity has now a component along the gradient of the density. Vortices tend to spiral out from the high density region towards places where the gradient of the density vanishes: local minima of the density distribution. This behavior is depicted in Fig.\,\ref{fig:trajectory} which shows an orbit of a vortex that was initially in a high density region.

Intriguingly, for $\gamma \ll 1 $, orbits spiraling out of the high density region tend to a hexagonal path that is a semistable limit cycle.  From the velocity field orientation one can see that the actual stable points correspond to the vertices of the first  Brillouin zone, that is, a honeycomb lattice rotated by $45 ^{\circ}$ with respect to the semistable limit cycle.
In order to better understand these aspects,  we turn to  a simplified model. 


\subsection{Triangular toy model}
\label{sec:triangular}
The vortex velocity in Eq.\,\eqref{eq:v_vortex_approx} depends only on the matter density distribution and not directly on the interaction potential. Hence, assuming that there exists an interaction potential that produces a certain matter distribution, we can determine  the  corresponding vortex velocity. 

As a toy model for the supersolid  density modulation one can take
\be
\rho(\bm r) = \rho_0 \left( 1 +  C \sum_{i=1}^3 \cos (\bm q_i \cdot \bm r )\right)\,,
\label{eq:triangle_density}
\ee
where the vectors
\be
\bm q_1 = q (0,1)\,, \quad \bm q_2 = q (\sqrt{3}/2,-1/2)\,, \quad \bm q_3 = -q (\sqrt{3}/2,1/2)\,,
\ee
indicate the vertices of a 2D triangular lattice. Hereafter we will take $q=1$, meaning that distances are measured in units of $1/q$. The 
the density contrast is 
\be
\frac{\rho_\textbf{max}-\rho_\textbf{min}}{\rho_\textbf{max}+\rho_\textbf{min}} =\frac{9 C}{4 + 3 C}\,,
\ee
and thus  $0\leq C \leq 2/3$. For definiteness, we take $C=0.3$, meaning that the density ranges between $0.7 \rho_0$ and  $1.3 \rho_0$. Such matter density distribution is smoother than the one produced by the model with  dipole-dipole and s-wave  interactions. However, it has the same  space symmetry and given its simplicity it is possible to be easily analyzed. 

\begin{figure}[b!]
     \centering
\includegraphics[width=0.45\textwidth]{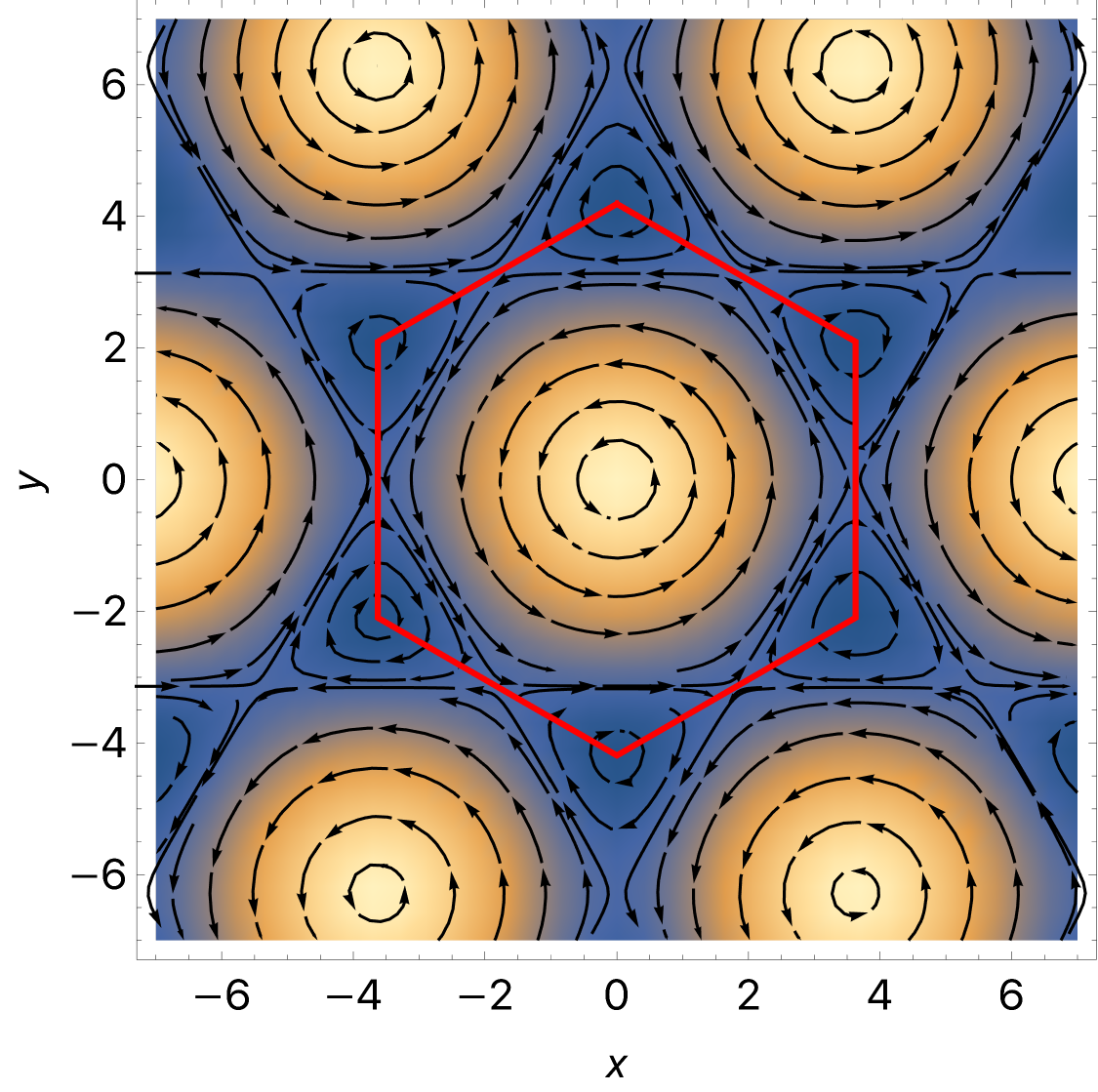}\,\,
\includegraphics[width=0.52\textwidth]{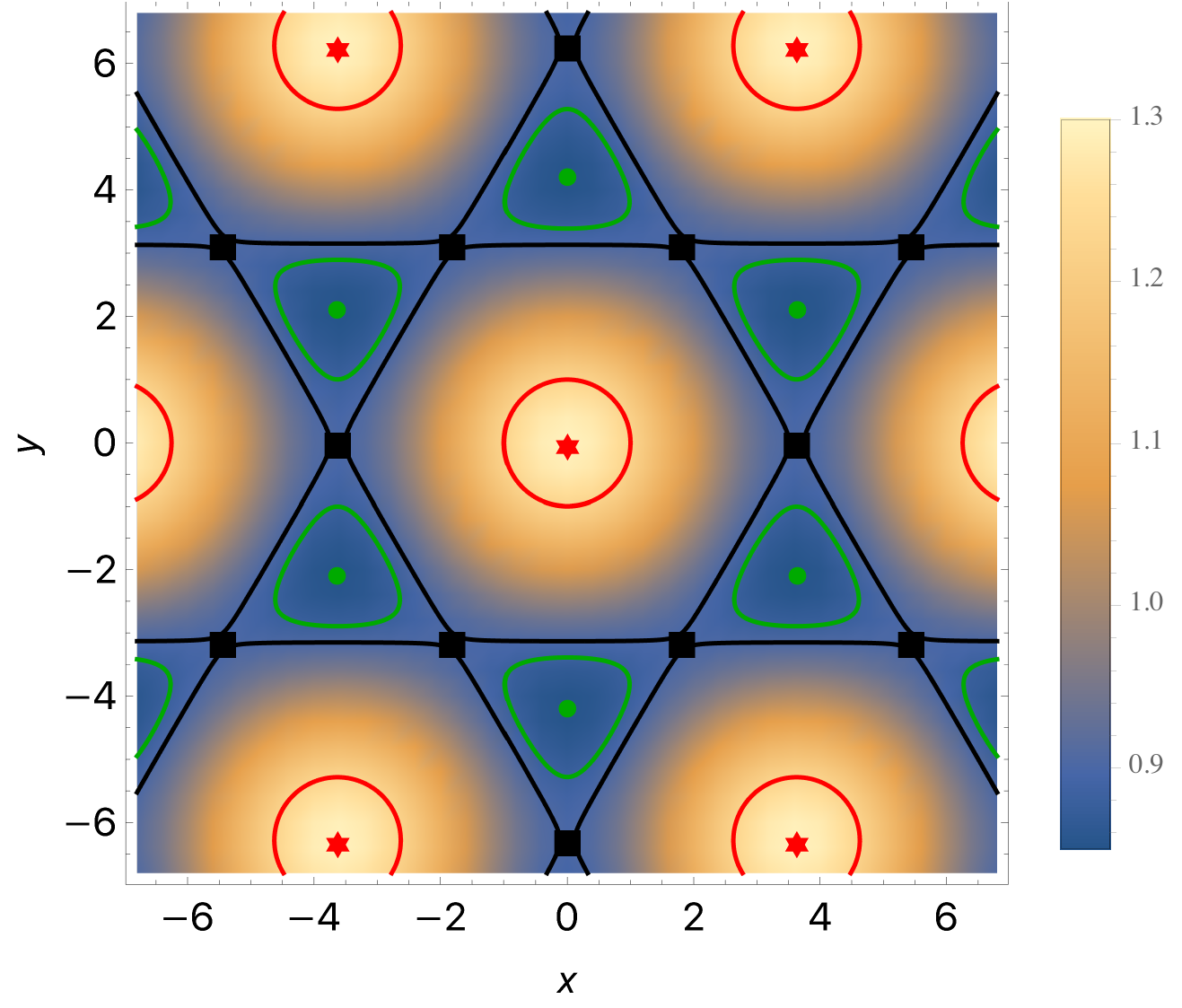}
     \caption{Triangular model, see Eq.\,\eqref{eq:triangle_density}, with $C=0.3$, in the case of vanishing dissipation. The number density is measured in units of $\rho_0$, distances in units of $1/q$.  Left: streamlines (in black) and first Brillouin zone (in red). Right: stationary points and orbits. Red stars correspond to maxima of the number density, while  green dots are minima. In both cases, orbits around these points are periodic. Filled black squares correspond to density saddle points. Orbits passing through a saddle point form a hexagonal path. A vortex that moves along such a path terminates at a saddle point.  }
     \label{fig:triangular}
 \end{figure}

The left panel of Fig.\,\ref{fig:triangular} shows the density plot in units of $\rho_0$,
the first Brillouin zone (in red) of the central lattice point and
 streamlines (in black) for vanishing dissipation.  The first Brillouin zone is a hexagon with vertices at the density minima. As in the case of the supersolid discussed above, since for $\gamma=0$ the system is conservative, streamlines form  closed paths.

If we indicate with $\bm x_v = (x,y)$ the vortex position,  for vanishing dissipation we obtain an analytical  first integral  given by
\be
y(x) = \pm \arccos\left\{  -\frac{1}{2} \cos\left(\frac{\sqrt{3}}{2} x\right) \pm \frac{1}{2 \sqrt{2}} \sqrt{2\left[1+2 \cos \left(\frac{y_0}2\right)\right]^2-1 + \cos(\sqrt{3}x) }\right\}\,,
\ee
where $y_0 \equiv y(x=0)$. The first integral is shown
in the right panel of Fig.\,\ref{fig:triangular} for different values of $y_0$.
For  $y_0 \ll 1$, we can linearize and obtain the circular  orbit 
\be
y^2 + x^2 = y_0^2\,,
\ee
around the density maxima (in red).
With increasing values of $y_0$ circles  are deformed until, for   $y_0=\pi$, the first integral is a hexagon centered at a red star. Such hexagons  enclose all the  periodic orbits centered at the density maximum and are rotated  $45 ^{\circ}$ with respect to the first Brillouin zone. Note that each  orbit starting from a point exactly on the hexagon terminates at one of the saddle points (in black). 
For   $\pi< y_0 < 4/3 \pi$, periodic orbits are instead around density minima (in green). From the right panel of Fig.\,\ref{fig:triangular}, one can see that the whole plane can  be tiled in hexagons and triangles. Hexagons are centered at density maxima, while triangles are instead centered at density minima.

\begin{figure}[b!]
     \centering
\includegraphics[width=0.45\textwidth]{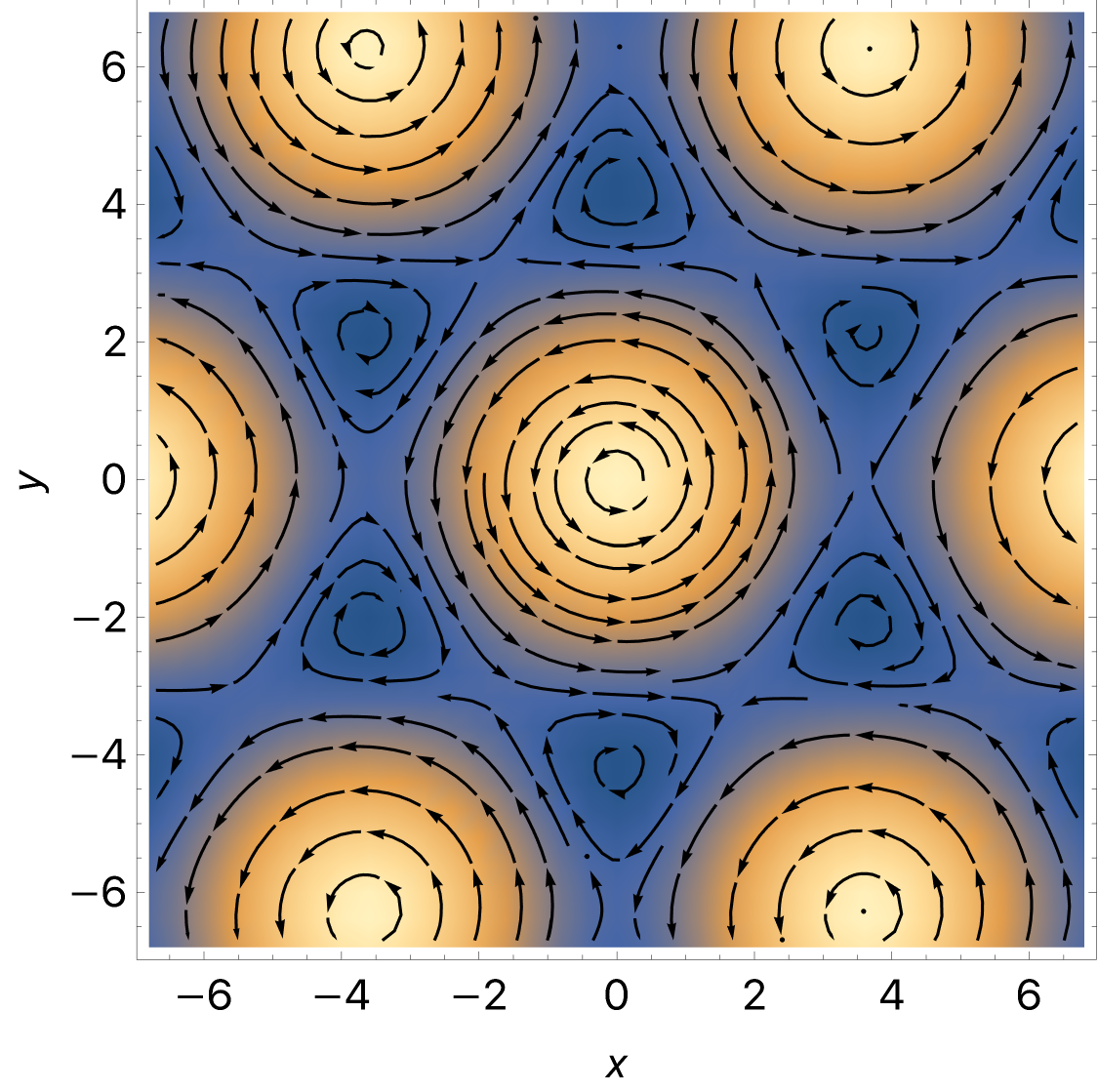}\,\,
\includegraphics[width=0.52\textwidth]{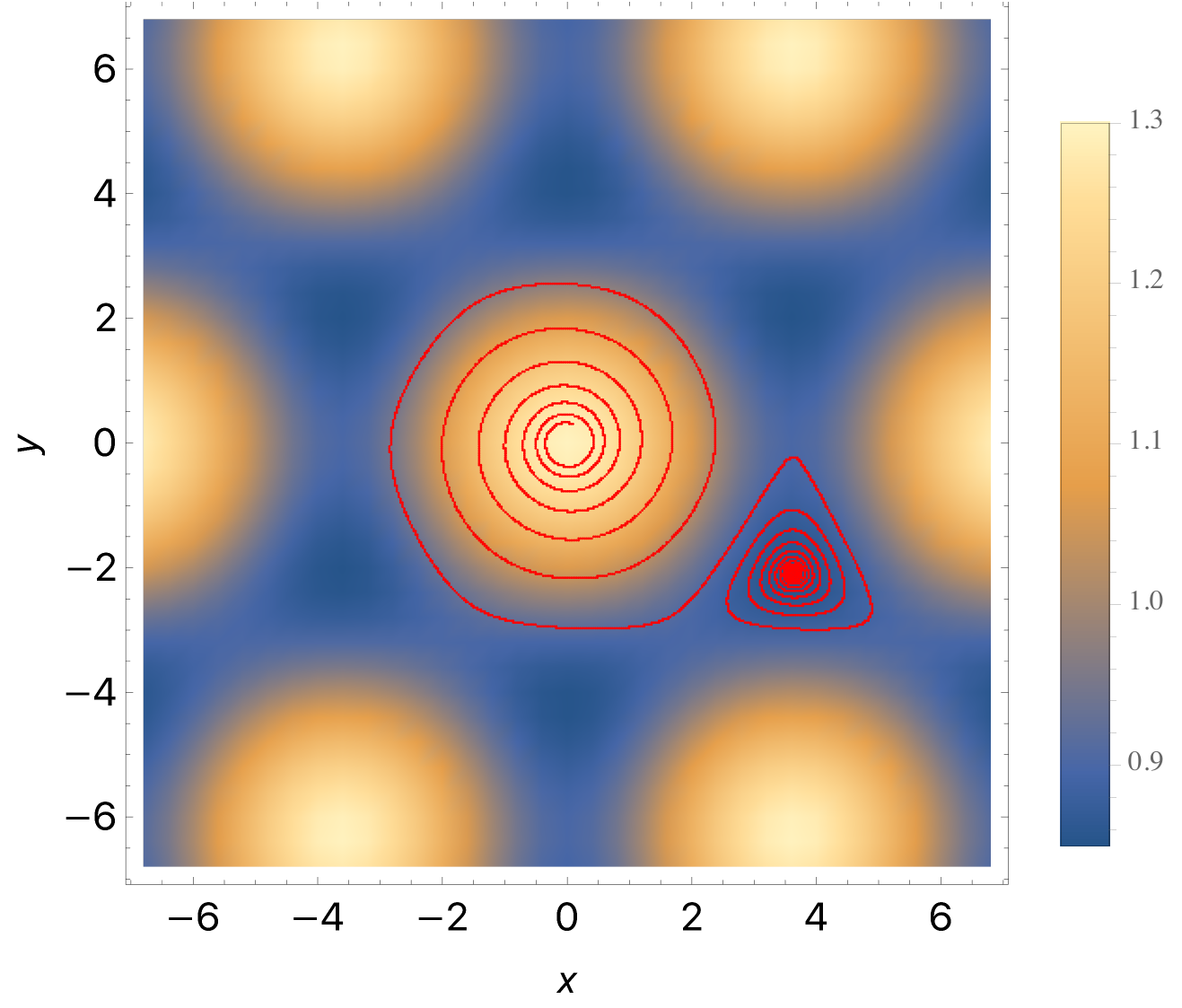}
     \caption{Triangular model, see Eq.\,\eqref{eq:triangle_density}, with $C=0.3$ and $\gamma=0.05$. The number density is measured in units of $\rho_0$, distances in units of $1/q$. Left: dissipation  deforms periodic orbits; streamlines  converge to density minima. Right: orbit of a vortex starting in the high density region. The vortex  first  spirals out the unstable point and it is then  attracted by an asymptotically stable focus.  }
     \label{fig:triangular_dissipative}
 \end{figure}

Fig.\,\ref{fig:triangular_dissipative} shows  the streamlines (left) as well as an example of orbit (right) including dissipation, with $\gamma=0.05$. 
As in  the case of supersolid described in the previous section, when $\gamma >0$, orbits are deformed. Maxima of the number density become unstable focus points, while  minima become asymptotically stable focus points. No periodic orbits exist: all orbits terminate in an asymptotically stable fixed point. In more detail, orbits starting in the high density region spiral out until  they reach the boundary of one of the triangles surrounding a green dot, shown in  the right panel of Fig.\,\ref{fig:triangular}. Then, vortices are captured by the corresponding attractor. The time needed to reach an attractor starting from a point in the high density region depends on $1/\gamma$. For  $\gamma \to 0$, such time diverges. 
A detailed analysis of the vortex time evolution  as well as of the basins of attraction is underway.

\section{Conclusion}\label{sec:conclusion}
 We have numerically simulated the inner crust of
  NSs  with a supersolid of  $^{164}$Dy atoms. 
 These two apparently different systems share several  key ingredients. They have a high density crystalline structure, made of neutron rich nuclei in NSs, and of clusters of bosons in the quantum gas.   A   superfluid bath embeds the crystalline structure, in NSs it is a neutron superfluid, while in supersolids it is a boson superfluid.
Finally, in both cases the periodic structure consists of superfluid matter, thus  both systems can be viewed as a supersolids.

  When a supersolid is put in rotation,  quantized vortices organize themselves between the crystal sites. We have provided an extended discussion of the 
  energy variation arising from the presence of vortices. 
  During an imposed braking process, vortices are pinned, and thus the superfluid component does not spin down. A glitch event happens when a vortex leaves its pinning site. 
  The pinning sites as well as the direction of vortex percolation can be inferred by analyzing the energy profile obtained imprinting a vortex on the top of the nonrotating system. Insight on the vortex evolution can be gained by a simple toy model with triangular symmetry. The appealing feature of such model is that in the non-dissipative case an exact first integral exists, which allows to better understand vortex orbits.

These findings establish a strong connection between quantum mechanics and astrophysics, offering a new perspective on the inner nature of NSs. A better understanding of glitches will provide valuable insights into the internal structure and dynamics of nuclear matter under extreme conditions, which cannot be directly realized through terrestrial laboratory experiments. Therefore, the existence of an analogous system on Earth, dipolar supersolids, presents an unparalleled opportunity to replicate and test the vortex and crystal dynamics inside NSs, opening new avenues for the quantum simulation of celestial bodies.

\backmatter


\bmhead{Acknowledgements}

This study received support from the European Research Council through the Advanced Grant DyMETEr (No.\,101054500), the QuantERA grant MAQS by the Austrian Science Fund (FWF) (No.\,I4391-N), the DFG/FWF via Dipolare E2 (No.\,I4317-N36) and a joint-project grant from the FWF (No.\,I4426). 
E.P.\,acknowledges support by the FWF within the DK-ALM (No. W1259-N27). T.B.\,acknowledges financial support by the ESQ Discovery programme (Erwin Schrödinger Center for Quantum Science \& Technology), hosted by the Austrian Academy of Sciences (ÖAW).


\end{document}